\documentclass[12pt]{article}

\usepackage{geometry}
\usepackage{graphicx}
\usepackage[utf8]{inputenc}
\usepackage{amsmath,amscd}    % need for subequations
\usepackage{graphicx}   % need for figures
\usepackage{verbatim}   % useful for program listings
\usepackage{color}      % use if color is used in text
\usepackage{subfigure}  % use for side-by-side figures
\usepackage{hyperref}   % use for hypertext links, including those to external documents and URLs
\usepackage{bbm}
\usepackage{amsfonts}
\usepackage{mathtools}
\usepackage{amssymb}
\usepackage{tensor}
\begin{document}

%\maketitle
\author{Niels G. Gresnigt}

\title{A topological model of composite preons from the minimal ideals of two Clifford algebras}
\maketitle

\abstract{

We demonstrate a direct correspondence between the basis states of the minimal ideals of the complex Clifford algebras $\mathbb{C}\ell(6)$ and $\mathbb{C}\ell(4)$, shown earlier to transform as a single generation of leptons and quarks under the Standard Model's unbroken $SU(3)_c\times U(1)_{em}$ and $SU(2)_L$ gauge symmetries respectively, and a simple topologically-based toy model in which leptons, quarks, and gauge bosons are represented as elements of the braid group $B_3$.  

It was previously shown that mapping the basis states of the minimal left ideals of $\mathbb{C}\ell(6)$ to specific braids replicates precisely the simple topological structure describing electrocolor symmetries in an existing topological preon model. This paper extends these results to incorporate the chiral weak symmetry by including a $\mathbb{C}\ell(4)$ algebra, and identifying the basis states of the minimal right ideals with simple braids. The braids corresponding to the charged vector bosons are determined, and it is demonstrated that weak interactions can be described via the composition of braids. 

}
%%%%%%%%%%%%%%%%%%%%%%%%%%%%%%%%%%%%%%%%%%%%%%%%
\section{Introduction}

Grand unified theories (GUT) and preon models represent two approaches to motivating the Standard Model's (SM) symmetry group, $SU(3)_c\times SU(2)_L\times U(1)_Y$, and particle content from more fundamental principles. The former approach merges the gauge groups of the SM into a single semi-simple Lie group. Such GUTs, including the famous $SU(5)$ and $\textrm{Spin}(10)$ theories \cite{georgi1974unity}, invariably predict additional gauge bosons, interactions, and proton decay, none of which have thus far been observed. Even if the unification of the SM gauge groups is successful, the choice of Lie group still requires justification, given the infinite possibilities. Preon models were developed with the hope of deriving the properties of the SM particles from a smaller set of constituent particles. The most famous of these is the Harari-Shupe preon model, based on just two fundamental particles \cite{Harari1979,Shupe1979}. 

Using the Harari-Shupe model as inspiration, it was shown in \cite{Bilson-Thompson2005} that one generation of SM fermions can be represented in terms of simple braids composed of three twisted ribbons and two crossings, connected together at the top and bottom to a parallel disk. In this model,the \textit{twist structure} of the ribbons accounts for the electrocolor symmetries, with charges of $\pm e/3$ represented by integral twists of $\pm 2\pi$ on the ribbons, and the permutations of the twisted ribbons representing color. The \textit{braid structure} of the ribbons on the other hand encodes the weak symmetry and chirality. The weak interaction is then represented topologically via braid composition. The twist and braid structure are not individually topologically conserved, but rather are interchangeable \cite{gresnigt2019combing,Bilson-Thompson2009}.

 %Mathematically speaking, elementary particles then correspond to elements of $(B_2)^3\rtimes (B_3^c)$, where the $(B_2)^3$ factor of the Artin braid group $B_2$ contains the twisting of the three ribbons (twist structure), and the circular Artin braid group $B_3^c$ contains their braiding (braid structure)\footnote{In the circular Artin braid group, the strands composing the braid are attached at top and bottom to a disk.} \cite{gresnigt2019combing,Bilson-Thompson2009}. The semi-direct product encodes the fact that the twisting or ribbons is permuted by the braiding. An adaptation of the braid model to the minimal $SU(5)$ GUT is given in \cite{cartin2015braids}. 

%%%%%%%%%%%%%%%%%%%%%%%%%%%%
There has recently also been some efforts to generate the SM symmetries and particle content from Clifford algebras, particularly those that arise when tensor products of division algebras act on themselves from the left or from the right\footnote{There are many related approaches that look at the division algebra as a basis for SM physics, see for example  \cite{gunaydin1973quark,gunaydin1974quark,dixon2013division,furey2016standard,manogue2010octonions,burdik2018hurwitz,gresnigt2019braided,perelman2019r,gillard2019three}. }. The basis states of the minimal left ideals of the Clifford algebra $\mathbb{C}\ell(6)$, generated via the left adjoint actions of $\mathbb{C}\otimes\mathbb{O}$, transform precisely as a single generation of leptons and quarks under the electrocolor group $SU(3)_C\otimes U(1)_{EM}$\cite{furey2016standard}. Furthermore, the chiral weak symmetry can be described via the minimal right ideals of $\mathbb{C}\ell(4)$ \cite{furey20183}. In these models, the finite particle content in the model is derived from the basis states of the finite-dimensional minimal left and right ideals of Clifford algebras. These ideals are constructed from a Witt decomposition of the algebra, and the gauge symmetries then correspond to the unitary symmetries that preserve this Witt decomposition.

In \cite{gresnigt2018braids} (see also \cite{gresnigt2017braidsgroups}) it was shown that by identifying the basis states of the minimal left ideals of $\mathbb{C}\ell(6)$ with particular braids in the circular braid group $B_3^c$\footnote{In the circular Artin braid group, the strands composing the braid are attached at top and bottom to a disk.}, and subsequently exchanging the resulting braiding for twisting, that the twist structure responsible for the electrocolor symmetry in the braid model \cite{Bilson-Thompson2005} is replicated. This paper extends this result to include the chiral weak symmetry, by identifying the basis states of the minimal right ideals of $\mathbb{C}\ell(4)$ with suitable braids in $B_3$. The minimal ideal basis states then generate the appropriate braid structure, which differs from the original model \cite{Bilson-Thompson2005}, with it no longer being the case that all particles are represented by braids with two crossings. The weak force is represented topologically via braid composition.

\section{Standard Model particle states from the minimal ideals of Clifford algebras}
%%%%%%%%%%%%%%%%%%%%%%%%%%%%%%%%%%%%%%%%%%%%%%%
\subsection{Electro-color symmetries from $\mathbb{C}\ell(6)$}

In \cite{furey2016standard} it was shown that a Witt decomposition of the complex Clifford algebra $\mathbb{C}\ell(6)$ decomposes the algebra into two minimal left ideals whose basis states transform as a single generation of leptons and quarks under the unbroken electrocolor $SU(3)_c\times U(1)_{em}$. A Witt basis of $\mathbb{C}\ell(6)$ can be defined as\footnote{following the convention of \cite{furey2016standard}.}
\begin{eqnarray}
\alpha_1&\equiv&\frac{1}{2}(-e_5+ie_4),\qquad \alpha_2\equiv\frac{1}{2}(-e_3+ie_1),\qquad \alpha_3\equiv \frac{1}{2}(-e_6+ie_2),\\
\alpha_1^{\dagger}&\equiv& \frac{1}{2}(e_5+ie_4),\qquad \alpha_2^{\dagger}\equiv\frac{1}{2}(e_3+ie_1),\qquad \alpha_3^{\dagger}\equiv \frac{1}{2}(e_6+ie_2),
\end{eqnarray}
satisfying the anticommutator algebra of fermionic ladder operators
\begin{eqnarray}\label{mtis}
\left\lbrace \alpha_i^{\dagger},\alpha_j^{\dagger} \right\rbrace=\left\lbrace \alpha_i,\alpha_j \right\rbrace=0,\qquad \left\lbrace \alpha_i^{\dagger},\alpha_j \right\rbrace =\delta_{ij}.
\end{eqnarray}
From these nilpotents one can then construct the minimal left ideal $S^u\equiv \mathbb{C}\ell(6)\omega\omega^{\dagger}$, where $\omega\omega^{\dagger}=\alpha_1\alpha_2\alpha_3\alpha_3^{\dagger}\alpha_2^{\dagger}\alpha_1^{\dagger}$ is a primitive idempotent. Explicitly:
\begin{eqnarray}\label{Ideal1}
\nonumber S^u\equiv &{}&
\;\;\nu \omega\omega^{\dagger}+\\
\nonumber \bar{d}^r\alpha_1^{\dagger}\omega\omega^{\dagger} &+& \bar{d}^g\alpha_2^{\dagger}\omega\omega^{\dagger} + \bar{d}^b\alpha_3^{\dagger}\omega\omega^{\dagger}\\
\nonumber u^r\alpha_3^{\dagger}\alpha_2^{\dagger}\omega\omega^{\dagger} &+& u^g\alpha_1^{\dagger}\alpha_3^{\dagger}\omega\omega^{\dagger} + u^b\alpha_2^{\dagger}\alpha_1^{\dagger}\omega\omega^{\dagger}\\
&+& e^{+}\alpha_3^{\dagger}\alpha_2^{\dagger}\alpha_1^{\dagger}\omega\omega^{\dagger},
\end{eqnarray}
where $\nu$, $\bar{d}^r$ etc. are suggestively labeled complex coefficients denoting the isospin-up elementary fermions. The conjugate system gives a second, linearly independent, minimal left ideal of isospin-down fermions $S^d\equiv \mathbb{C}\ell(6)\omega^{\dagger}\omega$ spanned by the states
\begin{eqnarray}
\left\lbrace 1,\alpha_1, \alpha_2,\alpha_3, \alpha_2\alpha_3,\alpha_3\alpha_1, \alpha_1\alpha_2,\alpha_1\alpha_2\alpha_3\right\rbrace\omega^{\dagger}\omega.
\end{eqnarray}
The Witt decomposition is preserved by the electrocolor symmetry $SU(3)_c\times U(1)_{em}$, with each basis state transforming as a specific lepton or quark as indicated by their suggestively labeled complex coefficients. The reader is directed to \cite{furey2016standard} for the explicit representation of the $SU(3)_c\times U(1)_{em}$ generators.

%%%%%%%%%%%%%%%%%%%%%%%%%%%%%%%%%%%%%%%%%%%%%
\subsection{Including weak symmetries via $\mathbb{C}\ell(4)$}
Including the $SU(2)_L$ weak symmetry requires an additional Clifford algebra, whose Witt decomposition is preserved by $SU(2)$. This algebra is $\mathbb{C}\ell(4)$ with a nilpotent basis $\{\beta_1,\beta_2,\beta_1^{\dagger},\beta_2^{\dagger}\}$ of ladder operators. Following the construction of minimal left ideals of $\mathbb{C}\ell(6)$, two four complex-dimensional minimal right ideals are given by $\Omega\Omega^{\dagger}\mathbb{C}\ell(4)$ and $\Omega^{\dagger}\Omega\mathbb{C}\ell(4)$, where $\Omega=\beta_2\beta_1$ and $\Omega^{\dagger}=\beta_1^{\dagger}\beta_2^{\dagger}$. Explicitly the ideals are spanned by the states
\begin{eqnarray}
\{\Omega^{\dagger}\Omega,\Omega^{\dagger}\Omega\beta_1^{\dagger},\Omega^{\dagger}\Omega\beta_2^{\dagger},\Omega^{\dagger}\Omega\beta_1^{\dagger}\beta_2^{\dagger}\},\;\{\Omega\Omega^{\dagger},\Omega\Omega^{\dagger}\beta_1,\Omega\Omega^{\dagger}\beta_2,\Omega\Omega^{\dagger}\beta_2\beta_1\}.
\end{eqnarray}

%%%%%%%%%%%%%%%%%%%%%%%%%%%%%%%%%%%%%%%%%%%%%%%%%%%%%%%%%%%%%%%%%%%%%%%%%%%%%%%%%%%%%%
\subsection{Combining $\mathbb{C}\ell(6)$ electrocolor and $\mathbb{C}\ell(4)$ weak states}

To fully describe weak transformations requires that we combine the $\mathbb{C}\ell(6)$ minimal left ideal states with the $\mathbb{C}\ell(4)$ minimal right ideal states. Each state then simultaneously belongs to a $\mathbb{C}\ell(6)$ minimal left ideal and a $\mathbb{C}\ell(4)$ minimal right ideal.

The neutrino $\nu$ is represented by the $\mathbb{C}\ell(6)$ minimal left ideal basis state $\omega\omega^{\dagger}$. Via the $\mathbb{C}\ell(4)$ right ideals, we can now include chirality, so that
\begin{eqnarray}
\nu_R=\omega\omega^{\dagger}\Omega^{\dagger}\Omega,\qquad
\nu_L=\omega\omega^{\dagger}\Omega^{\dagger}\Omega\beta_1^{\dagger}.
\end{eqnarray}
Similarly, the neutrino's weak doublet partner, the electron $e^-$ in its left- and right-handed form can now be written as
\begin{eqnarray}
e^-_L=\alpha_1\alpha_2\alpha_3\omega^{\dagger}\omega\Omega^{\dagger}\Omega\beta_2^{\dagger},\qquad
e^-_R=\alpha_1\alpha_2\alpha_3\omega^{\dagger}\omega\Omega^{\dagger}\Omega\beta_1^{\dagger}\beta_2^{\dagger}.
\end{eqnarray}
Notice that the neutrino and electron live in different $\mathbb{C}\ell(6)$ minimal left ideals, but in the same $\mathbb{C}\ell(4)$ minimal right ideal. One can write down the quark states in a similar manner (see \cite{gresnigt2020standard} for details). In summary, the eight weak-doublets are identified as
\begin{eqnarray}\label{weak states}
\begin{pmatrix} \nu_L \\ e^-_L  \end{pmatrix}=\begin{pmatrix} \omega\omega^{\dagger}\Omega^{\dagger}\Omega\beta_1^{\dagger} \\ 
\alpha_1\alpha_2\alpha_3\omega^{\dagger}\omega\Omega^{\dagger}\Omega\beta_2^{\dagger}\end{pmatrix},\qquad
\begin{pmatrix} u^{(3)}_L \\ d^{(3)}_L  \end{pmatrix}=\begin{pmatrix} \alpha_j^{\dagger}\alpha_i^{\dagger}\omega\omega^{\dagger}\Omega^{\dagger}\Omega\beta_1^{\dagger} \\ 
\epsilon_{ijk}\alpha_k\omega^{\dagger}\omega\Omega^{\dagger}\Omega\beta_2^{\dagger}\end{pmatrix},\\
\begin{pmatrix} e^+_R\\ \bar{\nu}_R \end{pmatrix}=\begin{pmatrix} \alpha_3^{\dagger}\alpha_2^{\dagger}\alpha_1^{\dagger}\omega\omega^{\dagger}\Omega\Omega^{\dagger}\beta_2 \\
\omega^{\dagger}\omega\Omega\Omega^{\dagger}\beta_1 \end{pmatrix},\qquad
\begin{pmatrix} \bar{d}^{(3)}_R\\ \bar{u}^{(3)}_R \end{pmatrix}=\begin{pmatrix} \alpha_i^{\dagger}\omega\omega^{\dagger}\Omega\Omega^{\dagger}\beta_2 \\
\epsilon_{ijk}\alpha_j\alpha_k\omega^{\dagger}\omega\Omega\Omega^{\dagger}\beta_1 \end{pmatrix}
\end{eqnarray}
All of the other physical states are weak singlets. The appropriate $SU(2)$ generators that transform the states correctly via the weak symmetry $SU(2)_L$ are now given by
\begin{eqnarray}\label{final SU(2)}
\nonumber T_1&\equiv& -\beta_1\beta_2^{\dagger}\omega-\beta_2\beta_1^{\dagger}\omega^{\dagger},\\
T_2&\equiv& i\beta_1\beta_2^{\dagger}\omega-i\beta_2\beta_1^{\dagger}\omega^{\dagger},\\
\nonumber T_3&\equiv& \beta_1\beta_2^{\dagger}\beta_2\beta_1^{\dagger}\omega\omega^{\dagger}-\beta_2\beta_1^{\dagger}\beta_1\beta_2^{\dagger}\omega^{\dagger}\omega.
\end{eqnarray}

%%%%%%%%%%%%%%%%%%%%%%%%%%%%%%%%%%%%%%%%%%%%%%%%%%%%%%%%%%%%%%%%%%%%%%
\section{Mapping minimal ideal basis states to braided matter states}

With one generation of fermions identified algebraically in terms of the minimal ideals  $\mathbb{C}\ell(6)$ and $\mathbb{C}\ell(4)$, we now wish to map these algebraic states to topological braid states. In the braid model \cite{Bilson-Thompson2005}, a red quark state $u^r$ is written as $[1,0,1]\sigma_1\sigma_2^{-1}$, where the vector $[1,0,1]$ denotes the twist structure; in this case $2\pi$ clockwise twists on the first and third ribbon, and $\sigma_1\sigma_2^{-1}$ is the braid structure. The antiparticle state $\bar{u}^r$ is written $\sigma_2\sigma_1^{-1}[-1,0,-1]$, with the braiding first followed by the twist structure. Via braid composition, the particle and anti-particle states annihilate. It is important to note that the twist structure is in general permuted by the braid structure, so that, for example, $[1,0,1]\sigma_1\sigma_2^{-1}=\sigma_1\sigma_2^{-1}[0,1,1]$.

We replicate this structure in three steps. First we replicate the twist structure by identifying the basis states of the minimal left ideals of $\mathbb{C}\ell(6)$ with specific braids in $B_3^c$ (as was done in \cite{gresnigt2018braids}). Second, we include braid structure by identifying the $\mathbb{C}\ell(4)$ ladder operators with suitable braids in $B_3$, in such a way that the weak force is represented topologically via braid composition. That is, the $\mathbb{C}\ell(6)$ ideals are responsible for generating twist structure, whereas the $\mathbb{C}\ell(4)$ ideals are responsible for generating braid structure. This results in all states being written as $[a,b,c]B$, with the twist structure first and the braid structure second. The final step is then to change the order of the twist and braid structure for half of the states. 

%%%%%%%%%%%%%%%%%%%%%%%%%%%%%%%%%%%%%%%%%%%
\subsection{Electro-color twist structure}

The key to replicating the twist structure from the $\mathbb{C}\ell(6)$ minimal ideals is the observation that the twist structure and braid structure in the braid model are not individually topological invariants, but rather are interchangeable \cite{Bilson-Thompson2009,gresnigt2019combing}. One finds that: 
\begin{eqnarray}
\nonumber [1,0,0]=\sigma_3\sigma_2[0,0,0],\quad [0,1,0]=\sigma_1\sigma_3[0,0,0],\quad [0,0,1]=\sigma_2\sigma_1[0,0,0],
\end{eqnarray}
and it follows that the twist structure of every state in $S^u$ can be written as a braid in $B_3^c$. If one now maps each $\alpha_i^{\dagger}$ to a product of two braid generators, 
\begin{eqnarray}\label{alphadagger}
\alpha_1^{\dagger}\mapsto(\sigma_3\sigma_2),\;\;\alpha_2^{\dagger}\mapsto(\sigma_1\sigma_3),\;\;\alpha_3^{\dagger}\mapsto(\sigma_2\sigma_1),
\end{eqnarray}
then each basis state in $S^u$ (see (\ref{Ideal1})) maps uniquely to the twist structure of a braid model state, in a one-to one manner. As an example, consider a green up quark $u^g$:
\begin{eqnarray}
u^g: \alpha_1^{\dagger}\alpha_3^{\dagger}\omega\omega^{\dagger}\mapsto (\sigma_3\sigma_2)(\sigma_2\sigma_1)[0,0,0]=[1,1,0].
\end{eqnarray}
We can do the same for the second ideal, $S^d$, via the maps
\begin{eqnarray}\label{alphas}
\alpha_1\mapsto(\sigma_2^{-1}\sigma_3^{-1}),\;\;\alpha_2\mapsto(\sigma_3^{-1}\sigma_1^{-1}),\;\;\alpha_3\mapsto(\sigma_1^{-1}\sigma_2^{-1}).
\end{eqnarray}
The braids associated with $\alpha_i^{\dagger}$ and $\alpha_i$ are simply braid inverses of one another\footnote{This is a slight but important deviation from the original construction in \cite{gresnigt2018braids}}. It is readily checked that $\omega\omega^{\dagger}[0,0,0]=\omega^{\dagger}\omega[0,0,0]\omega\omega^{\dagger}=[0,0,0]$, as well as  $\omega^{\dagger}[0,0,0]=[0,0,0]\omega^{\dagger}=[1,1,1]$, and $\omega[0,0,0]=[0,0,0]\omega=[-1,-1,-1]$. It follows that the twist structure of the braid model \cite{Bilson-Thompson2005}, responsible for the electrocolor symmetries, is recovered from the two minimal ideals $S^u$ and $S^d$. 
%%%%%%%%%%%%%%%%%%%%%%%%%%%%%%%%%%%%%%%%%%%%%%%%5
\subsection{Weak braid structure}

We next focus on the braid structure, and identify the $\mathbb{C}\ell(4)$ ladder operators with suitable braids in $B_3$, in such a way that the weak force is topologically represented via braid composition\footnote{$\mathbb{C}\ell(6)$ has a total of six ladder operators, and the circular braid group $B_3^c$ has six generators. $\mathbb{C}\ell(4)$ has four ladder operators and we therefore need a braid group with four generators. This is the braid group $B_3$. This is also consistent with the braid model \cite{Bilson-Thompson2005} where the braid structure is strictly in $B_3$ and not $B_3^c$.}. This means that we want to define $\beta_i^{\dagger}$ and $\beta_i$ in terms of $\sigma_1,\sigma_2,\sigma_1^{\-1}$, and $\sigma_2^{-1}$. This section contains the main results of this paper.

There seems to be no a priori reason to choose one braid structure over another. As with the $\mathbb{C}\ell(6)$ ladder operators, we would like $\beta_i$ and $\beta_i^{\dagger}$ to map to inverse braids. One choice would be to map $\beta_1^{\dagger}\mapsto \sigma_1\sigma_2$, and $\beta_2^{\dagger}\mapsto\sigma_2\sigma_1$, however in that case both $\alpha_3^{\dagger}$ and $\beta_2^{\dagger}$ map to the same braid product. Taking inspiration from the braid model we define the following maps
\begin{eqnarray}
\beta_1^{\dagger}&\mapsto& \sigma_1\sigma_2^{-1},\qquad \beta_2^{\dagger}\mapsto \sigma_1^{-1}\sigma_2,\\
\beta_1 &\mapsto& \sigma_2\sigma_1^{-1},\qquad \beta_2\mapsto \sigma_2^{-1}\sigma_1.
\end{eqnarray}
With this choice, $\Omega\Omega^{\dagger}=\Omega^{\dagger}\Omega=\mathbb{I}$ where $\mathbb{I}$ represents the identity (unbraid). 

Subsequently, every physical states in $\mathbb{C}\ell(6)\otimes\mathbb{C}\ell(4)$ can now be mapped to a specific braid with both twist structure and braid structure. For example, consider a weak doublet consisting of $u^r_L$ and $d^r_L$. We have
\begin{eqnarray}
\begin{pmatrix} u^{r}_L \\ d^{r}_L  \end{pmatrix}=\begin{pmatrix} \alpha_3^{\dagger}\alpha_2^{\dagger}\omega\omega^{\dagger}\Omega^{\dagger}\Omega\beta_1^{\dagger} \\ 
\alpha_1\omega^{\dagger}\omega\Omega^{\dagger}\Omega\beta_2^{\dagger}\end{pmatrix}\mapsto 
\begin{pmatrix} [1,0,1]\sigma_1\sigma_2^{-1} \\ 
[0,-1,0]\sigma_1^{-1}\sigma_2\end{pmatrix}=\begin{pmatrix} [1,0,1]\sigma_1\sigma_2^{-1} \\ 
\sigma_1^{-1}\sigma_2[-1,0,0]\end{pmatrix}.
\end{eqnarray}
It is here that it is important to remember that the twist structure is permuted by the braid structure, so that $[0,-1,0]\sigma_1^{-1}\sigma_2=\sigma_1^{-1}\sigma_2[-1,0,0]$. Similarly, for the associated weak singlets we find
\begin{eqnarray}
\begin{pmatrix} u^{r}_R   \end{pmatrix}&=&\begin{pmatrix} \alpha_3^{\dagger}\alpha_2^{\dagger}\omega\omega^{\dagger}\Omega^{\dagger}\Omega  \end{pmatrix}\mapsto\begin{pmatrix} [1,0,1]\mathbb{I}  \end{pmatrix},\\
\begin{pmatrix} d^{r}_R   \end{pmatrix}&=&\begin{pmatrix} \alpha_1\omega^{\dagger}\omega \Omega^{\dagger}\Omega\beta_1^{\dagger}\beta_2^{\dagger} \end{pmatrix}\mapsto\begin{pmatrix} \sigma_1\sigma_2^{-1}\sigma_1^{-1}\sigma_2[0,0,-1] \end{pmatrix}.
\end{eqnarray}
The full list of particle states are listed in the Appendix.

A couple of important deviations from the original model \cite{Bilson-Thompson2005} are apparent. First, it is no longer true that the braid structure of all particles is the same length. However it remains true that all weakly interacting particle states have the same length, and are equivalent to their representations in \cite{Bilson-Thompson2005}. Secondly, the right-handed neutrino and left-handed anti-neutrino both correspond to the unbraid, or equivalently, the vacuum
\begin{eqnarray}
\nu_R\omega\omega^{\dagger}\Omega^{\dagger}\Omega\mapsto [0,0,0]\mathbb{I},\qquad \bar{\nu}_L\omega^{\dagger}\omega\Omega\Omega^{\dagger}\mapsto \mathbb{I}[0,0,0].
\end{eqnarray}

%%%%%%%%%%%%%%%%%%%%%%%%%%%%%%%%%%%%%%%%%%%
\subsection{Charged vector bosons}

Finally we want to identify the  charged vector bosons $W^+$ and $W^-$ with specific braids in such a way that the weak interaction may be topologically represented via braid composition. We can find the relevant braid representations of the $W^+$ and $W^-$ bosons by looking at the $SU(2)_L$ generators (\ref{final SU(2)}). The key point is that charged vector bosons must change both the braid and the charge structure. This was the motivation behind the construction of \cite{gresnigt2020standard} where the $SU(2)_L$ generators are defined such that the change in electric charge in weak transformations is apparent. Algebraically,
\begin{eqnarray}
\nonumber\left[ T_1,u^r_L\right] &=&-u^r_LT_1\\
\nonumber&=&u^r_L(\beta_1\beta_2^{\dagger}\omega+\beta_2\beta_1^{\dagger}\omega^{\dagger}),\\
\nonumber &=& (\alpha_3^{\dagger}\alpha_2^{\dagger}\omega\omega^{\dagger}\Omega^{\dagger}\Omega\beta_1^{\dagger})(\beta_1\beta_2^{\dagger}\omega)\\
\nonumber &=& (\alpha_3^{\dagger}\alpha_2^{\dagger}\omega)\omega^{\dagger}\omega\Omega^{\dagger}\Omega\beta_2^{\dagger}\\
&=&\alpha_1\omega^{\dagger}\omega\Omega^{\dagger}\Omega\beta_2^{\dagger}=d^r_L,
\end{eqnarray}
where we have used the fact that $\omega$ (and similarly $\omega^{\dagger}$) commutes with the $\mathbb{C}\ell(4)$ ladder operators\footnote{See \cite{Bilson-Thompson2009,gresnigt2019combing}}. 

The $W^-$ and $W^+$ bosons may then be identified as
\begin{eqnarray}
W^-&=&\beta_1\beta_2^{\dagger}\omega\mapsto \sigma_2\sigma_1^{-1}\sigma_1^{-1}\sigma_2[-1,-1,-1],\\
W^+&=&\beta_2\beta_1^{\dagger}\omega^{\dagger}\mapsto \sigma_2^{-1}\sigma_1\sigma_1\sigma_2^{-1}[1,1,1],
\end{eqnarray}
and it follows that $W^+W^-=W^-W^+=\mathbb{I}[0,0,0]$. We can then represent the weak process $d^r_L\rightarrow u^r_LW^-$ as
\begin{eqnarray}
\nonumber d^r_L \rightarrow u^r_LW^- &\mapsto& \sigma_1^{-1}\sigma_2[-1,0,0],\\
\nonumber&=& \sigma_1^{-1}\sigma_2[0,1,1][-1,-1,-1],\\
&=&([1,0,1]\sigma_1\sigma_2^{-1})(\sigma_2\sigma_1^{-1}\sigma_1^{-1}\sigma_2[-1,-1,-1]),
\end{eqnarray}
and subsequently $W^-\rightarrow \bar{\nu}_Re^-_L$ as
\begin{eqnarray}
\nonumber W^-\rightarrow \bar{\nu}_Re^-_L &\mapsto& \sigma_2\sigma_1^{-1}\sigma_1^{-1}\sigma_2[-1,-1,-1]\\
&=&([0,0,0]\sigma_2\sigma_1^{-1})(\sigma_1^{-1}\sigma_2[-1,-1,-1]).
\end{eqnarray}

Algebraically, the weak force is automatically chiral. In terms of braids we have
\begin{eqnarray}
\nonumber u^r_RW^-&\mapsto& [1,0,1]\mathbb{I}\sigma_2\sigma_1^{-1}\sigma_1^{-1}\sigma_2[-1,-1,-1]\\
&=&\sigma_2\sigma_1^{-1}\sigma_1^{-1}\sigma_2[0,-1,0],\\
\nonumber d^r_RW^+&\mapsto& (\sigma_1\sigma_2^{-1}\sigma_1^{-1}\sigma_2[0,0,-1])(\sigma_2^{-1}\sigma_1\sigma_1\sigma_2^{-1}[1,1,1])\\
&=&[1,1,0]\sigma_1\sigma_2^{-1}\sigma_1\sigma_2^{-1}.
\end{eqnarray}
In neither case does the right-handed particle transform into a physical state, because at the algebra level, the transformed states do not live in a $\mathbb{C}\ell(4)$ minimal ideal.
%%%%%%%%%%%%%%%%%%%%%%%%%%%%%%%%%%%%%%%%%%%
\section{Discussion}

We have demonstrated that there exists a direct correspondence between the basis states of the minimal ideals of the complex Clifford algebras $\mathbb{C}\ell(6)$ and $\mathbb{C}\ell(4)$, and a simple topologically-based toy model in which leptons, quarks, and gauge bosons are represented as simple braids composed of three ribbons. The basis states of the minimal left ideals of $\mathbb{C}\ell(6)$ and minimal right ideals of $\mathbb{C}\ell(4)$ were previously shown to transform as a generation of leptons and quarks under the $SU(3)_c\times U(1)_{em}$ and $SU(2)_L$ gauge symmetries respectively \cite{gresnigt2020standard,furey20183}.  

The twist structure, which topologically encodes the electrocolor symmetries of leptons and quarks in the braid model \cite{Bilson-Thompson2005} is replicated from the minimal left ideals of $\mathbb{C}\ell(6)$. Each ladder operator obtained from a Witt decomposition of $\mathbb{C}\ell(6)$, and subsequently each basis state of the minimal left ideals $S^u$ and $S^d$, is identified with a simple braid in the circular braid group $B_3^c$, after which the resulting braiding is exchanged for twisting. This exchange of braiding for twisting is only possible in $B_3^c$ \cite{gresnigt2019combing,Bilson-Thompson2009}.

The braid structure, which topologically encodes the weak symmetry and chirality of leptons and quarks was then replicated from the minimal right ideal structure of $\mathbb{C}\ell(4)$. The ladder operators arising from a Witt decomposition of the algebra are again identified with simple braids, but this time in the braid group $B_3$ which has four generators, compared to six for $B_3^c$. Finally, the braids corresponding to charged vector bosons were determined and it was demonstrated what weak interactions can be described via the composition of braids.

The braid structure in our model differs slightly from that of the original model \cite{Bilson-Thompson2005}.  In the original model, the braid structure of all leptons and quarks consisted of two braid generators. This is no longer the case in the present model, however it remains true that all weakly interacting particles retain this feature. Furthermore, the braid representation of the charged vector bosons is no longer trivial. Together, these two new features of our model (which are dictated from the underlying algebraic structure of the minimal right ideals of $\mathbb{C}\ell(4)$) explain the chiral nature of the weak force in the braid model, something which the original model did not address.

A final interesting observation is that both the right-handed neutrino and left-handed anti-neutrino both correspond to the unbraid, or equivalently, the vacuum.

%The identification of charged vector bosons is different to the braid model, where all the bosons correspond to the unbraid but carrying non-trivial twist structure. In fact, in the model $W^-\mapsto \omega$ and $W^+\mapsto \omega^{\dagger}$. This identification however is insufficient to accurately represent weak interactions because there is nothing specifying that the gauge bosons act only on left-handed particles.

%For the anti-doublet, the minus sign indicates that the boson now acts from the left, or equivalently, that the $\mathbb{C}\ell(4)$ ladder operators are written in reverse order. 
 %%%%%%%%%%%%%%%%%%%%%%%%%%%%%%%%%%%%%%%%%%%%%%%%
\subsubsection*{Acknowledgments}

This work is supported by the Natural Science Foundation of the Jiangsu Higher Education Institutions of China Programme grant 19KJB140018 and XJTLU REF-18-02-03 grant.

\section*{Appendix}

\begin{center}
\begin{tabular}{ |c|c|c|c| } 
 \hline
 $\nu_L$ & $[0,0,0]\sigma_1\sigma_2^{-1}$ & $\nu_R$ & $[0,0,0]\mathbb{I}$\\
 \hline 
 $e_L^-$ & $\sigma_1^{-1}\sigma_2[-1,-1,-1]$ & $e_R^-$ & $\sigma_1\sigma_2^{-1}\sigma_1^{-1}\sigma_2[-1,-1,-1]$\\ 
 \hline
 $u^r_L$ & $[1,0,1]\sigma_1\sigma_2^{-1}$ & $u^r_R$ & $[1,0,1]\mathbb{I}$ \\
  \hline
 $u^g_L$ & $[1,1,0]\sigma_1\sigma_2^{-1}$ & $u^g_R$ & $[1,1,0]\mathbb{I}$\\
 \hline
 $u^b_L$ & $[0,1,1]\sigma_1\sigma_2^{-1}$ & $u^b_R$ & $[0,1,1]\mathbb{I}$\\
 \hline
 $d^r_L$ & $\sigma_1^{-1}\sigma_2[-1,0,0]$ & $d^r_R$ & $\sigma_1\sigma_2^{-1}\sigma_1^{-1}\sigma_2[0,0,-1]$\\
 \hline
 $d^g_L$ & $\sigma_1^{-1}\sigma_2[0,-1,0]$ & $d^g_R$ & $\sigma_1\sigma_2^{-1}\sigma_1^{-1}\sigma_2[-1,0,0]$\\
 \hline
 $d^b_L$ & $\sigma_1^{-1}\sigma_2[0,0,-1]$ & $d^b_R$ & $\sigma_1\sigma_2^{-1}\sigma_1^{-1}\sigma_2[0,-1,0]$\\
  \hline
 $\bar{d}^r_R$ & $[1,0,0]\sigma_2^{-1}\sigma_1$ & $\bar{d}^r_L$ & $[1,0,0]\sigma_2^{-1}\sigma_1\sigma_2\sigma_1^{-1}$\\
 \hline 
  $\bar{d}^g_R$ & $[0,1,0]\sigma_2^{-1}\sigma_1$ & $\bar{d}^g_L$ & $[0,1,0]\sigma_2^{-1}\sigma_1\sigma_2\sigma_1^{-1}$\\ 
 \hline
  $\bar{d}^b_R$ & $[0,0,1]\sigma_2^{-1}\sigma_1$ & $\bar{d}^b_L$ & $[0,0,1]\sigma_2^{-1}\sigma_1\sigma_2\sigma_1^{-1}$\\
  \hline
 $\bar{u}^r_R$ & $\sigma_2\sigma_1^{-1}[-1,0,-1]$ &  $\bar{u}^r_L$ & $\mathbb{I}[-1,0,-1]$\\
 \hline
  $\bar{u}^g_R$ & $\sigma_2\sigma_1^{-1}[-1,-1,0]$ & $\bar{u}^g_L$ & $\mathbb{I}[-1,-1,0]$\\
 \hline
  $\bar{u}^b_R$ &$\sigma_2\sigma_1^{-1}[0,-1,-1]$ & $\bar{u}^b_L$ & $\mathbb{I}[0,-1,-1]$\\
 \hline
 $e^+_R$ & $[1,1,1]\sigma_2^{-1}\sigma_1$& $e^+_L$& $[1,1,1]\sigma_2^{-1}\sigma_1\sigma_2\sigma_1^{-1}$\\
 \hline
 $\bar{\nu}_R$ & $\sigma_2\sigma_1^{-1}[0,0,0]$ & $\bar{\nu}_L$ & $\mathbb{I}[0,0,0]$\\
 \hline
\end{tabular}
\end{center}

%\begin{thebibliography}{99}
%\section*{References}
%%%%%%%%%%%%%%%%%%%%%%%%%%%%%%%%%%%%%%%%%%%%%%%%%%%%%%%%%%%%%%%%%%%%%
\bibliography{NielsReferences}  % Replace xxx by your  usercode (no extension)
\bibliographystyle{unsrt}  
%%%%%%%%%%%%%%%%%%%%%%%%%%%%%%%%%%%%%%%%%%%%%%%%%%%%%%%%%%%%%%%%%%%%%

\end{document}